\documentclass{aa}
\usepackage{graphicx}
\begin{document}

\title{The VIMOS-VLT Deep Survey\thanks{Based on data obtained with
 the European Southern Observatory on Paranal, Chile.} \\
Galaxy luminosity function per morphological type up to $z=1.2$}

\author{
     O.~Ilbert\inst{1,7}
\and S.~Lauger \inst{1}
\and L.~Tresse\inst{1}
\and V.~Buat \inst{1}
\and S.~Arnouts\inst{1}
\and O.~Le F\`evre \inst{1}
\and D. Burgarella \inst{1}
\and E.~Zucca\inst{2}
\and S.~Bardelli\inst{2}
\and G.~Zamorani\inst{2} 
\and D.~Bottini\inst{3}
\and B.~Garilli\inst{3}
\and V.~Le~Brun\inst{1}
\and D.~Maccagni\inst{3}
\and J.-P.~Picat\inst{4}
\and R. Scaramella\inst{5}
\and M.~Scodeggio\inst{3}
\and G.~Vettolani\inst{5}
\and A.~Zanichelli\inst{5}
\and C.~Adami\inst{1}
\and M.~Arnaboldi\inst{6}
\and M.~Bolzonella \inst{7} 
\and A.~Cappi\inst{2}
\and S.~Charlot\inst{8,9}
\and T.~Contini\inst{4}
\and S.~Foucaud\inst{3}
\and P.~Franzetti\inst{3}
\and I.~Gavignaud \inst{4,12}
\and L.~Guzzo\inst{10}
\and A.~Iovino\inst{10}
\and H.J.~McCracken\inst{9,11}
\and B.~Marano\inst{7}  
\and C.~Marinoni\inst{1}
\and G.~Mathez\inst{4}
\and A.~Mazure\inst{1}
\and B.~Meneux\inst{1}
\and R.~Merighi\inst{2} 
\and S.~Paltani\inst{1}
\and R.~Pello\inst{4}
\and A.~Pollo\inst{10}
\and L.~Pozzetti\inst{2} 
\and M.~Radovich\inst{6}
\and M.~Bondi\inst{5}
\and A.~Bongiorno\inst{7}
\and G.~Busarello\inst{6}
\and P.~Ciliegi\inst{2}  
\and Y.~Mellier\inst{9,11}
\and P.~Merluzzi\inst{6}
\and V.~Ripepi\inst{6}
\and D.~Rizzo\inst{4}
}

\offprints{O.~Ilbert, e-mail: olivier.ilbert1@bo.astro.it}
   
\institute{ 
Laboratoire d'Astrophysique de Marseille (UMR 6110), CNRS-Universit\'e de Provence, B.P.8, 13376 Marseille C\'edex 12, France 
\and INAF-Osservatorio Astronomico di Bologna, via Ranzani 1, 40127 Bologna, Italy 
\and INAF-IASF, via Bassini 15, 20133 Milano, Italy  
\and Laboratoire d'Astrophysique de l'Observatoire Midi-Pyr\'en\'ees (UMR 5572), CNRS-Universit\'e Paul Sabatier, 14 avenue E. Belin, 31400 Toulouse, France 
\and INAF-IRA, via Gobetti 101, 40129 Bologna, Italy 
\and INAF-Osservatorio Astronomico di Capodimonte, via Moiariello 16, 80131 Napoli, Italy 
\and Universit\`a di Bologna, Dipartimento di Astronomia, via Ranzani 1, 40127 Bologna, Italy 
\and Max-Planck-Institut f\"ur Astrophysik, Karl-Schwarzschild-Str. 1, 85740 Garching bei M\"unchen, Germany 
\and Institut d'Astrophysique de Paris (UMR 7095), 98 bis Boulevard Arago, 75014 Paris, France 
\and INAF-Osservatorio Astronomico di Brera, via Brera 28, 20121 Milano, Italy 
\and Observatoire de Paris-LERMA, 61 avenue de l'Observatoire, 75014 Paris, France 
\and European Southern Observatory, Karl-Schwarzschild-Str. 2, 85748 Garching bei M\"unchen, Germany 
}

\date{Received ... / Accepted ... }

\abstract{We have computed the evolution of the rest-frame $B$-band
luminosity function (LF) for bulge and disk-dominated galaxies since
$z=1.2$.  We use a sample of 605 spectroscopic redshifts with
$I_{AB}\le 24$ in the Chandra Deep Field South from the VIMOS-VLT Deep
Survey, 3555 galaxies with photometric redshifts from the COMBO-17
multi-color data, coupled with multi-color HST/ACS images from the
Great Observatories Origin Deep Survey. We split the sample in bulge-
and disk-dominated populations on the basis of asymmetry and
concentration parameters measured in the rest-frame $B$-band. We find
that at $z=0.4-0.8$, the LF slope is significantly steeper for the
disk-dominated population ($\alpha=-1.19 \pm 0.07$) compared to the
bulge-dominated population ($\alpha=-0.53 \pm 0.13$). The LF of the
bulge-dominated population is composed of two distinct populations
separated in rest-frame color : 68\% of red $(B-I)_{AB}>0.9$ and
bright galaxies showing a strongly decreasing LF slope $\alpha=+0.55
\pm 0.21$, and 32\% of blue $(B-I)_{AB}<0.9$ and more compact galaxies
which populate the LF faint-end. We observe that red bulge-dominated
galaxies are already well in place at $z\simeq1$, but the volume
density of this population is increasing by a factor 2.7 between
$z\sim 1$ and $z\sim 0.6$. It may be related to the building-up of
massive elliptical galaxies in the hierarchical scenario. In addition,
we observe that the blue bulge-dominated population is dimming by 0.7
magnitude between $z\sim 1$ and $z\sim 0.6$. Galaxies in this faint
and more compact population could possibly be the progenitors of the
local dwarf spheroidal galaxies.

    }

\titlerunning{Galaxy luminosity function per morphological type up to $z=1.2$}
\authorrunning{Ilbert~O. et al.}
\maketitle

\section{Introduction}

A central issue to understand galaxy formation is to
study the building up of the Hubble sequence. One approach is to
measure the evolution in numbers and luminosity of
different galaxy types using the luminosity function (LF). 
To this effect, large samples of galaxies are required with a robust
estimate of distances, luminosities and morphological types.

At low as well as at high redshifts, most luminosity functions 
have been measured using
galaxy samples classified by spectral type
(e.g. \cite{Madgwick02}2002, \cite{deLapparent03}2003) or by
photometric type (e.g. \cite{Lilly95}1995, \cite{Wolf03}2003,
\cite{Zucca06}2006). The direct interpretation of these results in
the framework of a galaxy formation scenario is not straightforward since
galaxies move from one spectral class to another by a passive evolution of
their stellar population. Another way to define the galaxy type is to
define a morphological type from the measurement of the structural 
parameters of the galaxies from image analysis.
Even if star formation evolution could affect
galaxy morphologies, a morphological classification is less sensitive to
the star formation history than a classification based on the
spectral energy distribution. 
This classification is also more robust to follow similar galaxies at
different redshifts.

Unfortunately, it is difficult to assemble large samples of
morphologically classified galaxies with measured
spectroscopic redshifts for $z > 0.3$, as 
high resolution images are required to perform a morphological
classification and a large amount of telescope time is needed to 
measure spectroscopic redshifts. This is why the largest samples
of galaxies with both high resolution morphology and
spectroscopic redshifts are not
exceeding $\sim300-400$ spectroscopic redshifts (e.g. \cite{Brinchmann98}1998,
\cite{Cassata05}2005). Larger samples of galaxies are obtained
using the photometric redshifts method (e.g. \cite{Wolf05}2005,
\cite{Bell05}2005), but a major drawback of the photometric redshift
method is the difficulty to control the systematic uncertainties which
are affecting the redshift estimate and to quantify the impact of these
systematics on the LF estimate.

This paper presents a study of the evolution of the galaxy LF as a
function of morphological type. We use the spectroscopic redshifts
from the VIMOS VLT Deep Survey on the Chandra Deep Field South (CDFS;
\cite{LeFevre04}2004). The spectroscopic sample selected at $I_{AB}
\le 24$ is twice larger than previous spectroscopic samples at similar
redshifts. We consolidate the results using 3555 photometric redshifts
estimated from the COMBO-17 multi-color data. We check that
photometric redshifts are not creating a systematic bias in the LF
measurement from a detailed comparison between photometric redshift and
spectroscopic redshift results. The morphological classification is
performed using the multi-color images of the Hubble Space
Telescope-Advanced Camera for Surveys released by the Great
Observatories Origin Deep Survey (\cite{Giavalisco04}2004). Different
methods to perform the morphological classification of this sample are
presented in the companion paper \cite{Lauger06}(2006) and compared
with a visual classification of the sample. A single wavelength
rest-frame morphological classification can be applied over the whole
redshift range. \cite{Lauger06}(2006) show that this classification
can separate two robust classes : bulge- and disk-dominated
galaxies. This paper presents the measurement of the LF evolution per
morphological type, based on this single wavelength rest-frame
morphological classification.

We use a flat lambda ($\Omega_m~=~0.3$, $\Omega_\Lambda~=~0.7$)
cosmology with $h~=~H_{\rm0}/100$~km~s$^{-1}$~Mpc$^{-1}$. Magnitudes
are given in the $AB$ system.


\section{Data set description}

We use the high-resolution images provided by the Great Observatories
Origin Deep Survey (GOODS, \cite{Giavalisco04}2004) on the Chandra
Deep Field South (CDFS) to perform the study of the galaxy
morphology. The images have been acquired with the Hubble Space
Telescope-Advanced Camera for Surveys (HST/ACS). The field covers 160
arcmin$^2$ and has been observed in four bands F435W, F606W, F775W,
F850LP (noted hereafter $B$, $V$, $i$, $z$ respectively).

We use the spectroscopic redshifts from the VIMOS VLT Deep Survey
(VVDS) on the CDFS (\cite{LeFevre04}2004). The spectroscopic
observations have been conducted with the multi-object spectrograph
VIMOS on the VLT-ESO Melipal. The spectroscopic targets are selected
on a pure magnitude criterion $17.5 \le I_{AB} \le 24$ from the ESO
Imaging Survey (\cite{Arnouts01}2001). The sample used in this paper
is limited to the area covered by GOODS. Our sample contains 670
objects (605 galaxies, 60 stars, 5 QSOs) with a secure measurement of
the redshift (confidence level greater than 80\%) and a mean redshift
of $0.76$.

Multi-color data from COMBO-17 are available on the CDFS field
(\cite{Wolf04}2004). These data consist in 12 medium-band filters and
5 broad-band filters from 3500$\rm{\AA}$ to 9300\AA. We also use the
near-infrared $J$ and $K$ band data (21600\AA) from the ESO Imaging Survey
(\cite{Arnouts01}2001).

\section{Photometric redshifts with {\it Le Phare}}

We apply the code {\it Le
Phare}\footnote{www.lam.oamp.fr/arnouts/LE\_PHARE.html} (S. Arnouts \&
O. Ilbert) on the COMBO-17 multi-color data completed by the NIR data
from EIS to compute photometric redshifts for the complete CDFS
sample. The photometric redshifts are measured with a standard
$\chi^2$ from the best fit template on multi-color data
(\cite{Arnouts02}2002). Our set of templates is composed of four
observed spectra from \cite{Coleman80}(1980) and one starburst SED
from GISSEL (\cite{Bruzual03}2003). These 5 templates have
been interpolated to increase the accuracy of the redshift estimate.

Our photometric redshift code significantly improves the
standard $\chi^2$ method (\cite{ilbert}2006). 
We compute the average difference in each band between
the observed apparent magnitudes and the predicted apparent magnitudes
derived from the best fit template for a restricted sample of 67
bright galaxies ($I_{AB}<20$) with a spectroscopic redshift. These
differences never exceed 0.2 and have an average value over all
filters of 0.06 magnitudes. We correct the predicted apparent magnitudes from
these systematic differences. This method of calibration corrects for
the small uncertainties existing in the filter transmission curves or
in the calibration of the photometric zero-points.

\begin{figure}
\centering
\includegraphics[height=9cm]{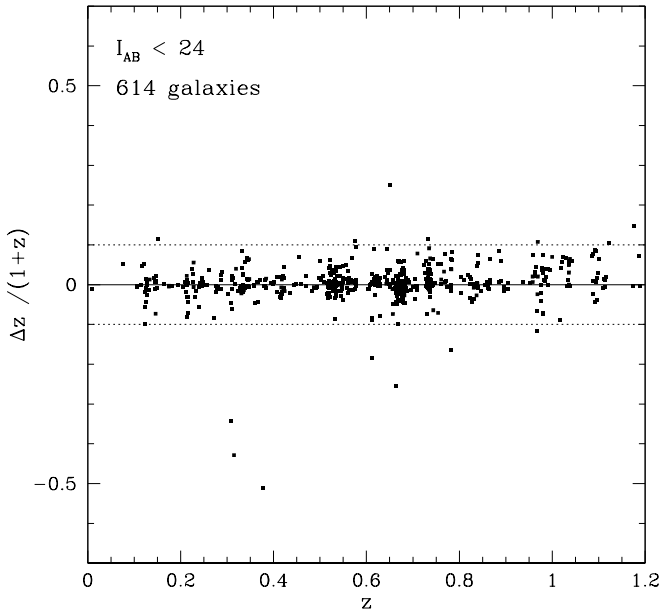}
\caption{Difference between VVDS spectroscopic redshifts and photometric
  redshifts ($\Delta z$) as a function of the spectroscopic redshift.}
\label{zphot}
\end{figure}

The comparison between photometric redshifts and spectroscopic
redshifts is shown in Fig.\ref{zphot}. We only use the most secure
spectroscopic redshifts with a confidence level greater than 95\%
(\cite{lefevre}2005). The
fraction of catastrophic errors ($\Delta z/(1+z)>0.1$) in 
photometric redshifts estimates is 1.1\% and the accuracy of the
measurement is $\sigma_{\Delta z/(1+z)}=0.046$. 295 stars are removed
from the sample which satisfy simultaneously a morphological criterion
(SExtractor CLASS\_STAR parameter greater than 0.975) and a $\chi^2$
criterion ($\chi^2(gal)-\chi^2(star)>0$). Up to $z=1.2$ and for
$I_{AB} \le 24$, the sample contains 3555 photometric redshifts of
galaxies associated with ACS/HST images.

\section{Morphological classification}

We measure asymmetry (A) and concentration (C) parameters on the
ACS/HST images (\cite{Abraham96}1996, \cite{Conselice00}2000,
\cite{Lauger05}2005, \cite{Menanteau06}2006). The concentration of
light is defined as the ratio between the radii which contain 80\% and
20\% of the total flux of the galaxy, respectively. 
The asymmetry is obtained by
computing the difference pixel per pixel of the original image and of
its 180$^\circ$ rotation. We adopt these parameters to define our
morphological classes since this classification is automatic,
quantitative and reproducible.

Importantly, and 
thanks to the multi-color coverage of the ACS/HST images, we can
measure the parameters $A$ and $C$ in the same rest-frame $B$-band
from $z=0$ up to $z\sim 1.2$ (\cite{Cassata05}2005). In this way, we
reduce the effect of a morphological k-correction which gives a more
patchy appearance to the galaxies at higher redshift when observations
are restricted to one band (e.g. \cite{Kuchinski00}2000,
\cite{Burgarella01}2001).

\begin{figure}
\centering \includegraphics[height=9cm]{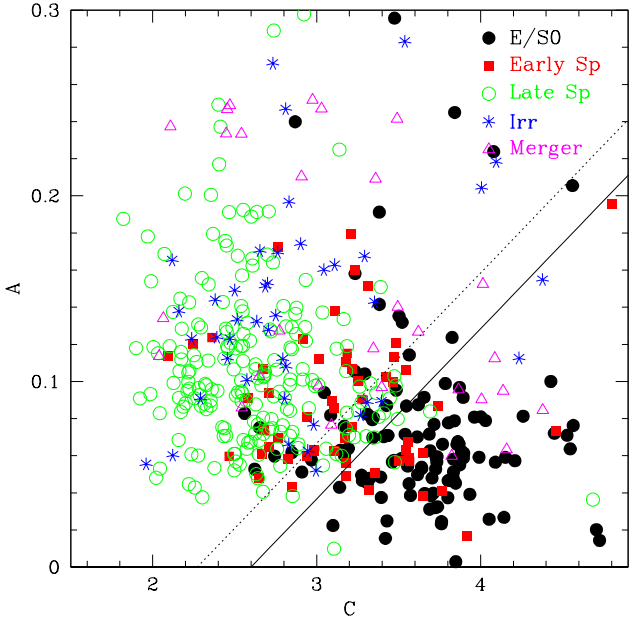}
\caption{Distribution in the A-C diagram of the eyeball classified
  galaxies with a spectroscopic redshift. The solid circles correspond
  to galaxies visually classified as elliptical-S0, the solid squares
  to early spirals, the open circles to late spirals, the star to
  irregulars and open triangles to mergers. The solid line is the
  empirical criterion $A = 0.0917 C-0.2383$ we have adopted to
  separate bulge- and disk-dominated populations. The dotted line
  corresponds to the criterion $A = 0.0917 C-0.2083$.}
\label{eyes}
\end{figure}

To relate the quantitative parameters A-C to the Hubble sequence, we
calibrate our morphological classes in the A-C plane with a visual
classification of galaxies. We adopt the empirical criterion $A=0.0917
C-0.2383$ to separate the bulge-dominated population from the
disk-dominated population (solid line of Fig.\ref{eyes}). This
criterion is chosen to maximize the separation between E/S0 and
spiral/irregular classified galaxies. We show in the companion paper
\cite{Lauger06}(2006) that the bulge-dominated population defined 
with $A\le 0.0917 C-0.2383$ contains  8.9\% of late spiral/irregular
galaxies and that the disk-dominated population includes
8.3\% of E/S0 galaxies down to $I_{AB}\le 24$. 
The bulge-dominated area contains
also $21.2\%$ of early spiral galaxies but the visual differentiation
between faint early spiral galaxies and lenticular galaxies is
strongly subjective.

\section{Galaxy luminosity function with $ALF$}

We derive the LF using the Algorithm for Luminosity Function (ALF)
described in \cite{Ilbert05}(2005). ALF includes 4 standard estimators
of the LF which are the 1/V$_{\rm max}$, C$^{+}$, SWML and
STY. Combining these 4 estimators allows us to check the robustness of
our estimate against spatial inhomogeneities, absolute magnitude
binning, or spectral type incompleteness (\cite{Ilbert04}2004). We
measure k-corrections from a procedure of template fitting on the
multi-color data (\cite{Ilbert05}2005), using either the
photometric redshift or the spectroscopic redshift according to
the sample used.

We measure the LF in the rest-frame $B$ Johnson band. At the average
redshift $z\sim0.76$ of this $I$-selected sample, this choice of the
rest-frame $B$-band limits the model dependency of the absolute
magnitudes (\cite{Ilbert05}2005), and minimizes any 
possible biases due to the mix
of spectral types (\cite{Ilbert04}2004).

\section{Results}

The values given in this section are obtained using the photometric
redshift sample to increase the accuracy of the measurements. However,
we systematically check that the results obtained with spectroscopic
redshifts and photometric redshifts are fully consistent.

\subsection{A blue bulge-dominated population}

\begin{figure}
\centering
\includegraphics[width=9cm]{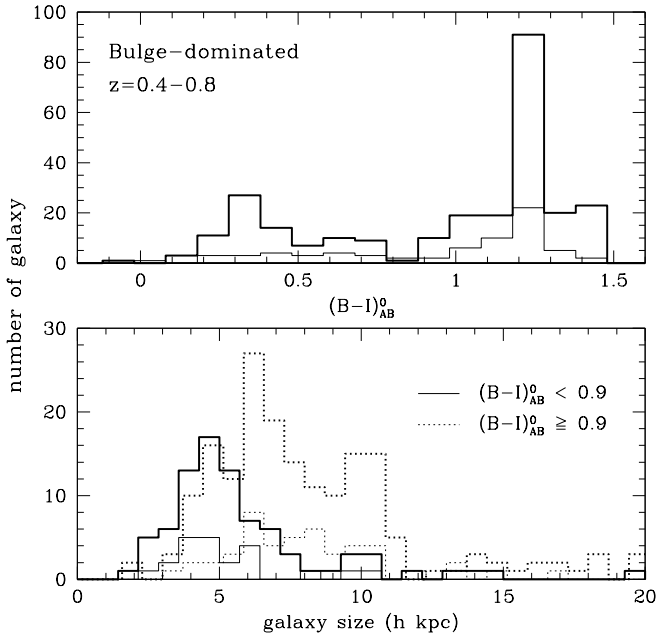}
\caption{Top panel: the distribution of the $(B-I)_{AB}^0$ rest-frame
colors for the bulge-dominated population. Bottom panel: galaxy size
distribution for the red bulge-dominated population (dotted line) and
for the blue bulge-dominated population (solid line). For both panels,
the thick lines correspond to the photometric redshift sample and the
thin lines to the spectroscopic redshift sample.}
\label{blue}
\end{figure}

 We observe a blue population of bulge-dominated galaxies as already
observed by e.g. \cite{Im01}(2001), \cite{Menanteau04}(2004),
\cite{Cross04}(2004). For both the spectroscopic and 
photometric redshift samples, the $(B-I)_{AB}^0$ rest-frame colors of
the bulge-dominated population present a bimodal distribution (see
upper panel of Fig.\ref{blue}). We split the bulge-dominated
population into two sub-samples separated in rest-frame color by
$(B-I)_{AB}^0=0.9$ according to this bimodality. We measure
32\%/68\% of blue/red galaxies at $z=0.4-0.8$, respectively. This
fraction of blue bulge-dominated galaxies is similar to the proportion
of 30\% obtained by \cite{Cross04}(2004) using a rest-frame
color criterion $(U-V)_{AB}>1.7$ and a similar selection of
$I_{AB}\le 24$.

To investigate farther the structural properties of this blue
population, we use the Petrosian radius $r(\eta=0.2)$ (see
\cite{Lauger05}2005) and the angular distance to measure the
physical size of the galaxy. Fig.\ref{blue} (lower panel) shows the
galaxy size distribution for blue and red bulge-dominated
galaxies. The blue population, with an average size of $5.8 \; h\;
kpc$ is more compact than the red population with an average size of
$8.2 \; h\; kpc$. The hypothesis that these two samples are
extracted from the same population is rejected at 99.9\% by a
Kolmogorov-Smirnov test.

\subsection{Shape of the LF versus morphology}

We investigate the dependency of the LF shape on the morphological
type. This analysis is performed in the redshift bin $0.4-0.8$, a
good compromise maximizing the number of galaxies and covering a
large absolute magnitude range. The LFs of the disk- and
bulge-dominated populations are shown in the middle panels of
Fig.\ref{LFevol} and the corresponding Schechter parameters are given
in Table~\ref{table1}. The LFs obtained with the photometric and the
spectroscopic redshift samples are in very good agreement and no
systematic trend in the shape is observed when using photometric
redshifts.

The LF of the disk-dominated population presents a steep slope
($\alpha=-1.19 \pm 0.07$) which contrasts with the decreasing slope
measured for the bulge-dominated population ($\alpha=-0.53 \pm
0.13$). This population of disk-dominated galaxies is the dominant
population of galaxies at $z\sim 0.6$. From the integration
of the LF up to $M_{B_{AB}}-5 log(h)=-17$, the disk-dominated
population represents 74\% of the whole galaxy population.

The slope measured for the bulge-dominated population
($\alpha=-0.53\pm 0.13$, see Table~\ref{table1}) is steeper than
previous LF measurement based on spectral type measurements
(e.g. $\alpha=0.52\pm 0.20$ for \cite{Wolf03}2003, $\alpha=-0.27 \pm
0.10$ for \cite{Zucca06}2006). The presence of the faint blue
bulge-dominated population explains this difference. 
The blue bulge-dominated population is composed of
faint galaxies representing 67\% of the bulge-dominated
galaxies for $-19.5 < M_{B_{AB}}-5log(h)$ but only 7.1\% for
$M_{B_{AB}}-5log(h) < -19.5$. As the LF slope is extremely steep,
the proportion of the observed blue galaxies is strongly dependent on
the considered limit (here $M_{B_{AB}}-5log(h) \le -17$). On the
contrary, the red bulge-dominated population is composed of bright
galaxies and represents 92\% of the bulge-dominated population for
$M_{B_{AB}}-5log(h) \le -19.5$. The density of red bulge galaxies
decreases toward fainter magnitudes with a strongly decreasing LF
slope $\alpha=0.55\pm 21$. Between $-19.5 < M_{B_{AB}}-5log(h) \le
-17$, the red bulge-dominated population represents only 5.7\% of the
whole population. The shape of the LF is in agreement with the
measurement done by \cite{Cross04}(2004) for red E/S0 galaxies
($\alpha=0.35\pm0.59$, $M^*_{B_{AB}}-5log(h)=-19.8\pm0.5$ mag at
$0.5<z<0.75$). This selected sample of red bulge-dominated galaxies
corresponds to the classical E/S0 population, composed of red and bright
galaxies with a strongly decreasing LF slope. Blue and red
bulge-dominated populations clearly exhibit different properties and need
to be analysed separately.

\begin{table*}
\begin{center}
\begin{tabular}{l l l l l  } 
\hline      
             &        &                    &    \multicolumn{1}{c}{$M^*_{AB}(B)-5log(h)$}  &  \multicolumn{1}{c}{$\phi^*$}  \\
  \multicolumn{1}{l}{Type}      &   \multicolumn{1}{c}{Nb}   &       \multicolumn{1}{c}{$\alpha$}     &     \multicolumn{1}{c}{(mag)}                & \multicolumn{1}{c}{($\times10^{-3} h^3 Mpc^{-3}$)} \vspace{0.1cm} \\
\hline \vspace{-0.1cm}\\    

 disk        &   892 &  ~~~~-1.19$^{{\rm + 0.07}}_{{\rm - 0.07}}$ & ~~~~~-20.22$^{{\rm + 0.15}}_{{\rm - 0.15}}$ &  ~~~~~12.39$^{{\rm + 2.18}}_{{\rm - 2.01}}$ \\
 bulge       &   261 &  ~~~~-0.53$^{{\rm + 0.13}}_{{\rm - 0.13}}$ & ~~~~~-20.20$^{{\rm + 0.19}}_{{\rm - 0.20}}$ &  ~~~~~7.48$^{{\rm + 1.04}}_{{\rm - 1.11}}$ \\

 red-bulge   &   178 &  ~~~~+0.55$^{{\rm + 0.21}}_{{\rm - 0.21}}$ & ~~~~~-19.53$^{{\rm + 0.14}}_{{\rm - 0.15}}$ &  ~~~~~7.44$^{{\rm + 0.56}}_{{\rm - 0.56}}$ \\
 blue-bulge  &    83 &    ~~~~-2.00$^{{\rm      }}_{{\rm      }}$ & ~~~~~-20.95$^{{\rm + 0.63}}_{{\rm - 0.79}}$ &  ~~~~~0.16$^{{\rm + 0.16}}_{{\rm - 0.09}}$ \\

\vspace{-0.2cm}\\   
\hline      
            
\end{tabular}
\caption{Schechter parameters for the rest-frame $B$-band LF in the
  redshift bin $0.4-0.8$ and the corresponding $1 \sigma$ error. The
  parameters without associated errors are fixed. Values are given for
  the photometric redshift sample.}
\label{table1}
\end{center}
\end{table*}

\subsection{Evolution of the LF}

\begin{figure*}
\centering \includegraphics[width=18cm]{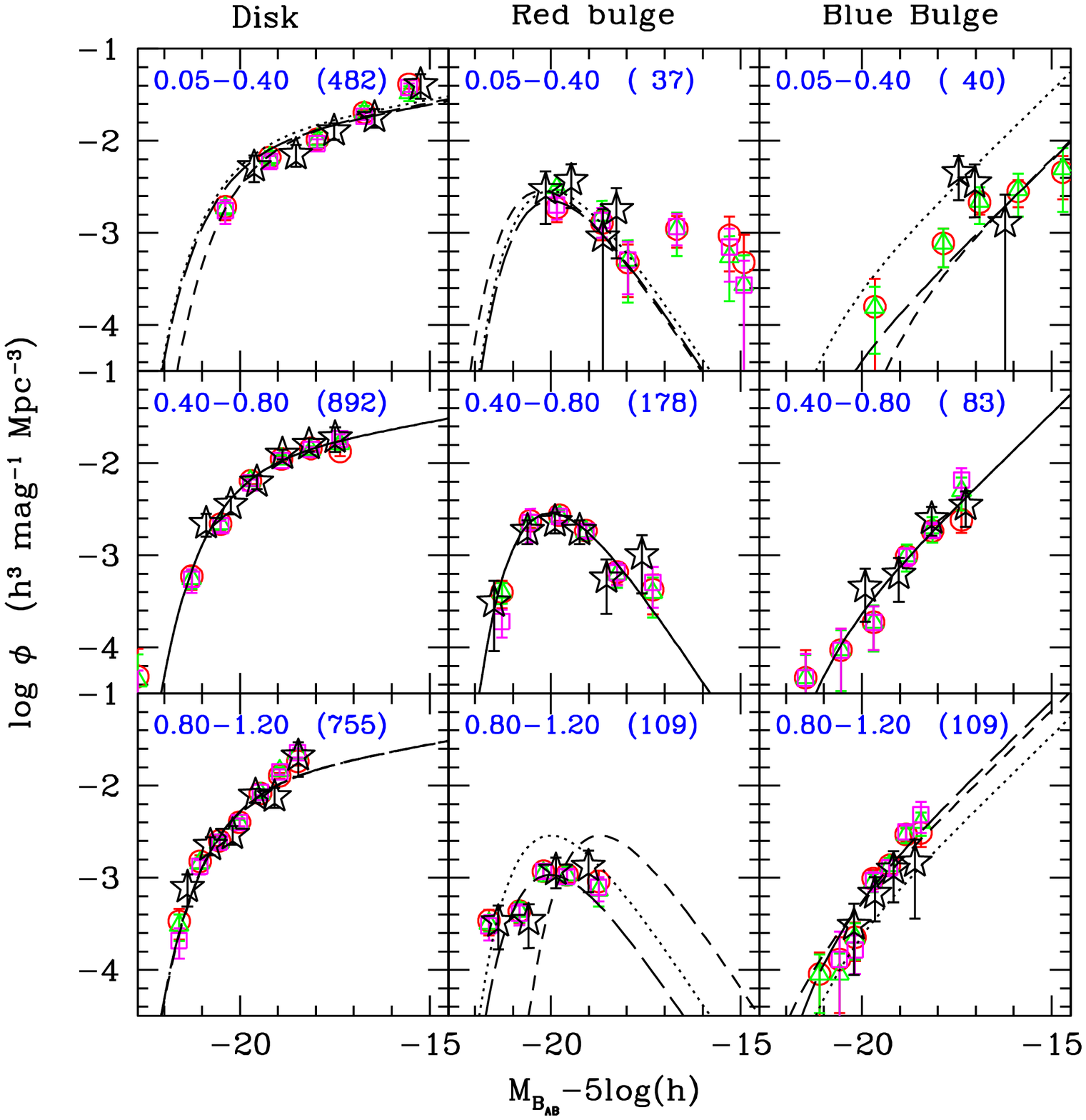}
\caption{Evolution of the Luminosity Function for galaxies
  classified in morphological types from $z=0.05$ up
  to $z=1.2$. Open stars correspond to the LFs measured with the
  spectroscopic redshift sample using the 1/V$_{\rm max}$
  estimator. All the other results are obtained using the photometric
  redshift sample. The left panels correspond to the LF of the
  disk-dominated population, the middle panels to the red
  bulge-dominated population and the right panels to the blue
  bulge-dominated population. The redshift bin and the corresponding
  number of galaxies is indicated in each panel. We adopt the
  following symbols for the various estimators: circles for the
  1/V$_{\rm max}$, triangles for the SWML, squares for the C$^+$ and
  solid lines for the STY (only at $z=0.4-0.8$ where the LF shape is
  constrained). The STY fit derived in the redshift bin $0.4-0.8$ is
  reported in each panel with dotted lines. The long dashed lines and
  short dashed lines correspond respectively to the fit obtained
  setting $\alpha-M^*$ (pure density evolution) and $\alpha-\Phi^*$
  (pure luminosity evolution) at the values obtained at $z=0.4-0.8$.}
\label{LFevol}
\end{figure*}

Figure \ref{LFevol} presents the evolution of the disk-, red bulge- and
blue bulge-dominated populations from $z=0.05$ up to $z=1.2$. The
Schechter parameters are given in Table~\ref{table2}. Since the
LF slope is not constrained at $z>0.8$, we set $\alpha$ to the value
measured at $z=0.4-0.8$.

The results obtained with the photometric redshift sample are fully in
agreement with the results obtained with the spectroscopic redshift
sample (see the open stars of Fig.\ref{LFevol}). This comparison
gives confidence that our measurement of the LF evolution is not due to
systematic trends in photometric redshift estimates.

The LF of the disk-dominated galaxies evolves only mildly over the
redshift range up to $z=1.2$. 
The slope of the disk-dominated galaxies in our sample 
is comparable to the local values obtained by
\cite{Marinoni99}(1999) ($\alpha=-1.10 \pm 0.07$ for S-Im eyeball
classified galaxies) or by \cite{Nakamura03}(2003) ($\alpha=-1.12 \pm
0.18$ for S-Im type defined with concentration parameter). No
significant evolution of $\Phi^*$ is measured between $z=0.05$ and
$z=1.2$ when testing for a pure density evolution (setting the $\alpha-M^*$
parameters, see Table.2). We measure only a small brightening of 0.4
magnitude when testing for pure luminosity evolution (setting the values of
$\alpha-\Phi^*$). Leaving $M^*$ and $\phi^*$ free (setting only
the slope), we measure a brightening of $\Delta M* \sim 0.5$
magnitude with no significant evolution of $\Phi^*$ between $z=0.05$
and $z=1.2$.

The density of the red bulge-dominated population decreases at high
redshift $z=0.8-1.2$. We measure a decrease in density by a factor 2.7
between $z\sim 0.6$ and $z\sim 1$ for a pure density evolution
(setting the $M^*-\alpha$ parameters). A pure luminosity evolution of
the LF is not a good fit of the non-parametric data (see
Fig.\ref{LFevol}). Leaving $M^*$ and $\phi^*$ free (setting only
the slope), we measure a small brightening of 0.2 mag with a
decrease of $\Phi^*$ by a factor 2 between $z\sim 0.6$ and $z\sim
1$, consistent with a pure density evolution. Due to the difficulty
to sample the bright part of the red-bulge dominated LF at
$z=0.05-0.4$, no strong conclusions can be drawn in this low
redshift range. The evolution of the LF for the red bulge-dominated
population is opposite to the increase in density observed by
\cite{Cross04}(2004) between $z=0.5-0.75$ and $z=0.75-1$ but is in
agreement with the result obtained by \cite{Ferreras05}(2005) on the
same field. Our results are fully consistent
with the results from the VVDS based on spectroscopic
redshifts and spectral type classification (\cite{Zucca06}2006). 
The observed evolution of the red bulge-dominated
population in our analysis remains small in comparison to 
the strong decrease in
density of the elliptical galaxies measured by \cite{Wolf03}(2003)
using spectral type classification and photometric redshifts. 
Our results instead show that the population of E/S0
galaxies is already mostly in place at $z\sim 1$.

The LF slope of the blue bulge-dominated population remains extremely
steep in all the redshift bins. This population strongly evolves. To
quantify the LF evolution of this population, 
we first set the $M^*-\alpha$ parameters to
the $z=0.4-0.8$ values and look for an evolution in $\Phi^*$. We
measure an increase in density of a factor 2.4 between $z=0.4-0.8$ and
$z=0.8-1.2$. Using the same procedure, we set the $\alpha - \Phi^*$
parameters to the $z=0.4-0.8$ values and look for an evolution in
$M^*$. We measure a brightening of 0.7 magnitude between $z=0.4-0.8$
and $z=0.8-1.2$. The same trend is also present between $z=0.05-0.4$
and $z=0.4-0.8$. It is unlikely than a large fraction of this blue
bulge-dominated population contains misclassified spiral
galaxies at high redshifts
as the visual inspection of the UV rest-frame images at $z\sim 1$
do not show star formation in the disc of these galaxies. If
a large fraction of these blue bulge-dominated galaxies is
including misclassified spiral galaxies at $z<0.7$ as claimed by
\cite{Ferreras05}(2005), the observed evolution of the blue
bulge-dominated galaxies should even be stronger than what
we have reported here.

\begin{table*}
\begin{center}
\begin{tabular}{l l l l l l  } 
\hline      
             &           &        &                    &     \multicolumn{1}{c}{$M^*_{AB}(B)-5log(h)$}  &  \multicolumn{1}{c}{$\phi^*$}  \\
   \multicolumn{1}{l}{Type}      &     \multicolumn{1}{c}{z}     &   \multicolumn{1}{c}{Nb}   &       \multicolumn{1}{c}{$\alpha$}     &     \multicolumn{1}{c}{(mag)}                & \multicolumn{1}{c}{($\times10^{-3} h^3 Mpc^{-3}$)}  \vspace{0.1cm}\\
\hline \vspace{-0.1cm}\\    

  disk       &  0.05-0.40 &   482 &  ~~~~-1.19$^{{\rm      }}_{{\rm      }}$ & ~~~~~-19.79$^{{\rm + 0.12}}_{{\rm - 0.11}}$ &  ~~~~~12.39$^{{\rm      }}_{{\rm      }}$ \\
             &  0.05-0.40 &   482 &  ~~~~-1.19$^{{\rm      }}_{{\rm      }}$ & ~~~~~-20.22$^{{\rm      }}_{{\rm      }}$   &  ~~~~~10.95$^{{\rm + 0.58}}_{{\rm - 0.58}}$ \\
             &  0.05-0.40 &   482 &  ~~~~-1.19$^{{\rm      }}_{{\rm      }}$ & ~~~~~-19.54$^{{\rm + 0.14}}_{{\rm - 0.15}}$ &  ~~~~~15.85$^{{\rm + 0.91}}_{{\rm - 0.87}}$ \\
             &  0.40-0.80 &   892 &  ~~~~-1.19$^{{\rm + 0.07}}_{{\rm - 0.07}}$ & ~~~~~-20.22$^{{\rm + 0.15}}_{{\rm - 0.15}}$ & ~~~~~ 12.39$^{{\rm + 2.18}}_{{\rm - 2.01}}$ \\
             &  0.80-1.20 &   755 &  ~~~~-1.19$^{{\rm      }}_{{\rm      }}$ & ~~~~~-20.21$^{{\rm + 0.04}}_{{\rm - 0.04}}$ &  ~~~~~12.39$^{{\rm      }}_{{\rm      }}$ \\
             &  0.80-1.20 &   755 &  ~~~~-1.19$^{{\rm      }}_{{\rm      }}$ & ~~~~~-20.22$^{{\rm       }}_{{\rm       }}$ &  ~~~~~12.44$^{{\rm + 0.51}}_{{\rm - 0.51}}$ \\
             &  0.80-1.20 &   755 &  ~~~~-1.19$^{{\rm      }}_{{\rm      }}$ & ~~~~~-20.06$^{{\rm + 0.07}}_{{\rm + 0.07}}$ &  ~~~~~15.65$^{{\rm + 1.06}}_{{\rm - 0.98}}$ \\

 red-bulge   &  0.05-0.40 &    37 &   ~~~~+0.55$^{{\rm      }}_{{\rm      }}$ & ~~~~~-19.77$^{{\rm + 0.40}}_{{\rm - 0.39}}$ &  ~~~~~7.44$^{{\rm      }}_{{\rm      }}$ \\
             &  0.05-0.40 &    37 &   ~~~~+0.55$^{{\rm      }}_{{\rm      }}$ & ~~~~~-19.53$^{{\rm       }}_{{\rm       }}$ &  ~~~~~ 5.79$^{{\rm + 1.23}}_{{\rm - 1.23}}$ \\
             &  0.40-0.80 &   178 &   ~~~~+0.55$^{{\rm + 0.21}}_{{\rm - 0.21}}$ & ~~~~~-19.53$^{{\rm + 0.14}}_{{\rm - 0.15}}$ &   ~~~~~7.44$^{{\rm + 0.56}}_{{\rm - 0.56}}$ \\
             &  0.80-1.20 &   109 &   ~~~~+0.55$^{{\rm      }}_{{\rm      }}$ & ~~~~~-18.23$^{{\rm + 0.09}}_{{\rm - 0.09}}$ &   ~~~~~7.44$^{{\rm      }}_{{\rm      }}$ \\
             &  0.80-1.20 &   109 &   ~~~~+0.55$^{{\rm      }}_{{\rm      }}$ & ~~~~~-19.53$^{{\rm      }}_{{\rm      }}$   &   ~~~~~2.78$^{{\rm + 0.30}}_{{\rm - 0.30}}$ \\
             &  0.80-1.20 &   109 &   ~~~~+0.55$^{{\rm      }}_{{\rm      }}$ & ~~~~~-19.76$^{{\rm + 0.09}}_{{\rm - 0.09}}$ &   ~~~~~3.05$^{{\rm + 0.29}}_{{\rm - 0.29}}$ \\

 blue-bulge  &  0.05-0.40 &    40 &  ~~~~-2.00$^{{\rm      }}_{{\rm      }}$ & ~~~~~-20.95$^{{\rm      }}_{{\rm      }}$ &   ~~~~~0.03$^{{\rm + 0.01}}_{{\rm - 0.01}}$ \\
             &  0.05-0.40 &    40 &  ~~~~-2.00$^{{\rm      }}_{{\rm      }}$ & ~~~~~-19.11$^{{\rm + 0.22}}_{{\rm - 0.20}}$ &   ~~~~~0.16$^{{\rm      }}_{{\rm      }}$ \\
             &  0.40-0.80 &    83 &  ~~~~-2.00$^{{\rm      }}_{{\rm      }}$ & ~~~~~-20.95$^{{\rm + 0.63}}_{{\rm - 0.79}}$ &   ~~~~~0.16$^{{\rm + 0.16}}_{{\rm - 0.09}}$ \\
             &  0.80-1.20 &   109 &  ~~~~-2.00$^{{\rm      }}_{{\rm      }}$ & ~~~~~-20.95$^{{\rm      }}_{{\rm      }}$ &   ~~~~~0.38$^{{\rm + 0.04}}_{{\rm - 0.04}}$ \\
             &  0.80-1.20 &   109 &  ~~~~-2.00$^{{\rm      }}_{{\rm      }}$ & ~~~~~-21.63$^{{\rm + 0.11}}_{{\rm - 0.10}}$ &   ~~~~~0.16$^{{\rm      }}_{{\rm      }}$ \\

\vspace{-0.2cm}\\   
\hline      
            
\end{tabular}
\caption{Evolution of the Schechter parameters for the rest-frame
  $B$-band LF and associated $1 \sigma$ errors. Values are given for
  the photometric redshift sample. The parameters without
  associated errors are set using the values measured at
  $z=0.4-0.8$. At $z=0.05-0.4$ and $z=0.8-1.2$, we always provide the
  parameters for a pure luminosity and density evolution. When
  possible, we also provide the parameters for $M^*-\Phi^*$ let free
  to vary. When
  $\alpha$ is set, the STY errors on $M^*$ and $\Phi^*$ are
  underestimated.}
\label{table2}
\end{center}
\end{table*}

Since $A$ and $C$ are quantitative parameters and since the
visual classification is strongly subjective, the separation between
bulge- and disk-dominated galaxies in the $A-C$ plan is not a sharp
limit. The area enclosed between $A\le 0.0917 C-0.2383$ (solid line of
Fig.\ref{eyes}) and $A\le 0.0917 C-0.2083$ (dotted line of
Fig.\ref{eyes}) contains a mix of different visual types. To quantify
the impact of the adopted criterion on our conclusions, we recompute
the LFs using the criteria $A\le 0.0917 C-0.2083$ rather than $A\le
0.0917 C-0.2383$. This less conservative criterion increases the
contamination of blue late spiral galaxies in the bulge-dominated
area. As expected, the LF normalization of the blue bulge-dominated
galaxies increases by a factor 1.6-2. We have
specifically used a conservative criterion throughout this paper ($A\le 0.0917
C-0.2383$) to limit the fraction of late spiral/irregular galaxies in
the bulge-dominated area. We have checked that only 12\% of the blue-bulge
dominated galaxies have been visually classified as late spiral or
irregular galaxies. The LFs of the disk-dominated and of the red
bulge-dominated galaxies remain unchanged and are little sensitive to
the adopted criterion.

\section{Discussion and conclusions}

We derive the rest-frame $B$-band LF of galaxies classified in 
morphological types up to
$z=1.2$ in the CDFS using the VVDS spectroscopic sample of 605
galaxies, 3555 photometric redshifts from COMBO-17 multi-color data
and the HST/ACS images from GOODS. We define bulge- and disk-dominated
populations on the basis of A-C parameters measured in the rest-frame
$B$-band (\cite{Lauger06}2006). We show that the LF of the
bulge-dominated population is the combination of two populations: a
red and bright population making 68\% of the bulge-dominated sample, 
and a blue population of more compact galaxies for the
remaining 32\% of the population. 
We observe a strong dependency of the LF shape on the
morphological type. In the redshift range $0.4<z<0.8$, we measure 
a shallow slope $\alpha=+0.55 \pm 0.21$ for the red bulge-dominated population
while  the disk-dominated population shows 
a very steep slope $\alpha=-1.19 \pm 0.07$. The blue
bulge-dominated population dominates the faint-end of the
bulge-dominated LF. We emphasize that without morphological 
information, the blue
bulge-dominated population can not be separated from the late spectral
type population even if a composite fit of the LF (\cite{deLapparent03}2003) is
computed as an alternative when visual morphologies are not
available.

We observe a small evolution of the disk-population. This is 
unexpected as the irregular galaxies included in our disk-dominated
class are expected to evolve strongly (e.g. \cite{Brinchmann98}1998).
As our one-wavelength A-C parameters are not
efficient to isolate the irregular or merger galaxies from the late spiral
galaxies (see Fig.\ref{eyes}), this small evolution of the
disk-dominated population could possibly be explained  by cosmic variance
in this small field, by the fact that spiral galaxies dominate the population
of irregular galaxies, and by the fact that we are insensitive to
the morphological k-correction effect. 

We measure an increase in density of the red, bright (brighter than
$M^*$ at $z>0.8$), bulge-dominated population with the age of the
Universe. The observed evolution of the red bulge-dominated LF could
be interpreted as evidence for the build-up of massive elliptical
galaxies from merging and accretion of smaller galaxies in a
hierarchical scenario.  Our results indicate that the population of
E/S0 galaxies is already in place at $z \sim 1$, in agreement with
e.g. \cite{Lilly95}(1995), \cite{Conselice05}(2005),
\cite{Zucca06}(2006).  As the field used in this paper is 160
arcmin$^2$ and includes large structures (\cite{Adami05}2005), cosmic
variance could possibly play a role and affect our results.  To
investigate this point, we compare the global LF measured on the CDFS
field with the LF in the VVDS-0226-04 field (\cite{Ilbert05}2005)
which covers an area 10 times larger. Correcting the LF normalizations
in the CDFS field to match the normalization of the global LF in the
VVDS-0226-04 field, we find that the evolution, although less
pronounced, still shows an increase of the red bulge-dominated density
with the age of the Universe.  The effect of the density-environment
relation (e.g. \cite{Dressler97}1997) on the observed evolution is
difficult to assess. One approach to address this uncertainty is to
compute the combined LF per morphological type and environment. 

We also observe a very strong evolution of the blue
bulge-dominated population corresponding to a brightening of 0.7
magnitude (or an increase in density by a factor 2.4) between $z
\sim 0.6$ and $z \sim 1$. The nature of this population remains
unclear.  Mergers expected in the hierarchical scenario could create a
burst of star formation explaining the blue color of these
galaxies. We have observed some  signs of disruption for a
significant fraction of these galaxies. These galaxies could be also a
dwarf population undergoing a strong burst of star formation in the galaxy
core, which could be interpreted as a bulge component
(\cite{Im01}2001). The smaller size of this blue population in
comparison to the red one as well as the faint absolute magnitude
distribution of this population seem to favor this
hypothesis. Considering that the evolution of the blue bulge-dominated
population produces a fading in luminosity, we can relate this evolution to
the strong decrease of the star formation rate observed from $z\sim 1$. This
blue population could be the progenitor of the local population of
dwarf spheroidal galaxies undergoing strong star formation at $z\sim
1$. Another possibility to investigate is the presence of an AGN in
the galaxy nucleus as shown by \cite{Menanteau05}(2005) which could be
strongly related to the star formation activity.

An increase in survey areas  is clearly necessary 
to limit the cosmic variance effects on computing
the LFs of morphologically selected populations, as is 
on-going with next generation surveys (e.g. COSMOS).
The development of quantitative methods to better isolate
morphological galaxy classes, in particular merger and irregular
galaxies, and the combination of morphological and spectral
classifications, will also be necessary for further progress.

\begin{acknowledgements}
This research has been developed within the framework of the VVDS
consortium.\\
This work has been partially supported by the
CNRS-INSU and its Programme National de Cosmologie (France),
and by Italian Ministry (MIUR) grants
COFIN2000 (MM02037133) and COFIN2003 (num.2003020150).\\
The VLT-VIMOS observations have been carried out on guaranteed
time (GTO) allocated by the European Southern Observatory (ESO)
to the VIRMOS consortium, under a contractual agreement between the
Centre National de la Recherche Scientifique of France, heading
a consortium of French and Italian institutes, and ESO,
to design, manufacture and test the VIMOS instrument.
\end{acknowledgements}

\end{document}